\documentclass[twocolumn,showpacs,preprintnumbers,amsmath,amssymb,prb]{revtex4}

\usepackage{graphicx}
\usepackage{dcolumn}
\usepackage{bm}
\usepackage{tabularx}
\usepackage{amsmath}

\begin{document}

\title{Multi-gap nodeless superconductivity in iron selenide FeSe$_x$: evident from quasiparticle heat transport}

\author{J. K. Dong,$^1$ T. Y. Guan,$^1$ S. Y. Zhou,$^1$ X. Qiu,$^1$ \\ L. Ding,$^1$ C. Zhang,$^1$ U. Patel,$^2$ Z. L. Xiao,$^2$ S. Y. Li$^{1,*}$}

\affiliation{$^1$Department of Physics, Surface Physics Laboratory (National Key Laboratory), and Laboratory of Advanced Materials, Fudan University, Shanghai 200433, P. R. China\\
$^2$Department of Physics, Northern Illinois University, DeKalb,
Illinois 60115, USA}

\date{\today}

\begin{abstract}
The in-plane thermal conductivity $\kappa$ of the iron selenide
superconductor FeSe$_x$ ($T_c$ = 8.8 K) were measured down to 120 mK
and up to 14.5 T ($\simeq 3/4 H_{c2}$). In zero field, the residual
linear term $\kappa_0/T$ at $ T \rightarrow 0$ is only about 16
$\mu$W K$^{-2}$ cm$^{-1}$, less than 4\% of its normal state value.
Such a small $\kappa_0/T$ does not support the existence of nodes in
the superconducting gap. More importantly, the field dependence of
$\kappa_0/T$ in FeSe$_x$ is very similar to that in NbSe$_2$, a
typical multi-gap $s$-wave superconductor. We consider our data as
strong evidence for multi-gap nodeless superconductivity in
FeSe$_x$. This kind of superconducting gap structure may be generic
for all Fe-based superconductors.
\end{abstract}


\maketitle

Just as CuO-plane is the basic building block of high-$T_c$ cuprate
superconductors, the FeAs-layer is the basic structure of the newly
discovered FeAs-based high-$T_c$
superconductors.\cite{Kamihara,XHChen,GFChen,ZARen,RHLiu,MRotter,XCWang}
The FeAs-layer consists of a Fe square planar sheet tetrahedrally
coordinated by As. However, unlike the rigid CuO-plane in cuprates,
partial substitution of Fe by Co or Ni, or As by P within the
FeAs-layer can effectively induce
superconductivity.\cite{ASefat1,GHCao,ASefat2,LJLi,SJiang} In this
sense, the discovery of superconductivity in binary FeSe$_x$ ($T_c
\simeq$ 8 K) is of great interests, since it only contains the
superconducting FeSe-layer which has identical structure as
FeAs-layer, and the Se deficiency may cause the
superconductivity.\cite{FCHsu} More remarkably, the onset $T_c$ can
be enhanced to as high as 37 K for FeSe$_x$ under high
pressure,\cite{YMizuguchi,SMedvedev,SMargadonna} which further
implies that superconductivity in FeSe$_x$ may have the same
mechanism as in FeAs-based superconductors.

For this new family of high-$T_c$ superconductors, the pairing
symmetry of its superconducting gap is a key to understand the
mechanism of superconductivity. Extensive experimental and
theoretical work have been done to address this important issue for
FeAs-based superconductors (for a theoretical review, see Ref. 17;
for an experimental review, see Ref. 18). Although there is still no
consensus, more and more evidences point to multi-gap nodeless
superconductivity, possibly an unconventional $s^{\pm}$ paring
mediated by antiferromagnetic fluctuations.\cite{IMazin2} For the
prototype FeSe$_x$ superconductor, however, there were very few
experiments to study the superconducting gap structure. This is due
to its relatively lower $T_c$ and lack of sizable high-quality
single crystals.\cite{SBZhang,UPatel} The measurements of in-plane
magnetic penetration depth for polycrystal FeSe$_{0.85}$ are in
favor of anisotropic $s$-wave superconducting gap or two gaps ($s +
s$).\cite{RKhasanov} To clarify this important issue, more
experimental work are needed for FeSe$_x$ superconductor.

Low-temperature thermal conductivity measurement is a powerful tool
to study the superconducting gap structure.\cite{HShakeripour} The
field dependence of the residual thermal conductivity $\kappa_0/T$
for BaNi$_2$As$_2$ ($T_c$ = 0.7 K) is consistent with a dirty fully
gapped superconductivity.\cite{NKurita} For
Ba$_{1-x}$K$_x$Fe$_2$As$_2$ ($T_c \simeq$ 30 K) and
BaFe$_{1.9}$Ni$_{0.1}$As$_2$ ($T_c$ = 20.3 K), a negligible
$\kappa_0/T$ was found in zero field, indicating a full
superconducting gap.\cite{XGLuo,LDing} However, $\kappa(T)$ was only
measured in magnetic fields up to $H$ = 15 T ($\simeq$ 1/4
$H_{c_2}$), thus can not show clearly whether the superconductivity
has multi-gap character in FeAs-based
superconductors.\cite{XGLuo,LDing}

In this paper, we measure the thermal conductivity $\kappa$ of a
FeSe$_x$ single crystal with $T_c = 8.8$ K down to 120 mK and up to
14.5 T ($\simeq$ 3/4 $H_{c_2}$) to probe its superconducting gap
structure. In zero field, $\kappa_0/T$ is about 16 $\mu$W / K$^2$
cm, less than 4\% of its normal-state value. Such a small
$\kappa_0/T$ should not come from the nodal quasiparticle
contribution. It may simply come from the slight overestimation when
doing extrapolation, due to the lack of lower temperature data. The
field-dependence of $\kappa_0/T$ is very similar to that in
multi-gap $s$-wave superconductor NbSe$_2$. Based on our data, it is
evident that FeSe$_x$ is a multi-gap nodeless superconductor.

FeSe$_x$ single crystals with nominal formula FeSe$_{0.82}$ were
grown via a vapor self-transport method.\cite{UPatel} The $ab$-plane
dimensions of as-grown crystals ranges from a few hundred $\mu$m to
1 mm. Energy Dispersive of X-ray (EDX) microanalysis (Hitachi
S-4800) show that the actual Fe:Se ratio is very close to 1:1 in our
FeSe$_x$ single crystals. The nominal formula FeSe$_{0.82}$ was used
in the initial work by Hsu et al..\cite{FCHsu} However, the actual
superconducting phase was later determined to be
FeSe$_{0.99\pm0.02}$ in Ref. 27 and FeSe$_{0.974\pm0.005}$ in Ref.
28. Therefore the EDX result of our FeSe$_x$ single crystals is
consistent with these two later reports.

The ac magnetization was measured in a Quantum Design Physical
Property Measurement System (PPMS). An as-grown single crystal with
dimensions 1.0 $\times$ 0.40 mm$^2$ in the plane and 190 $\mu$m
thickness along the $c$-axis was selected for transport study.
Contacts were made directly on the sample surfaces with silver
paint, which were used for both resistivity and thermal conductivity
measurements. The typical contact resistance is a few ohms at room
temperature and 1.5 K, which is not as good as that on
Ba$_{1-x}$K$_x$Fe$_2$As$_2$ and BaFe$_{1.9}$Ni$_{0.1}$As$_2$ single
crystals.\cite{XGLuo,LDing} In-plane thermal conductivity was
measured in a dilution refrigerator using a standard
one-heater-two-thermometer steady-state technique. Due to the small
size of the sample and the non-ideal contacts, good thermalization
between sample and the two RuO$_2$ thermometers can only be achieved
down to 120 mK. Magnetic fields were applied along the $c$-axis and
perpendicular to the heat current. To ensure a homogeneous field
distribution in the sample, all fields were applied at temperature
above $T_c$.

\begin{figure}
\includegraphics[clip,width=6.5cm]{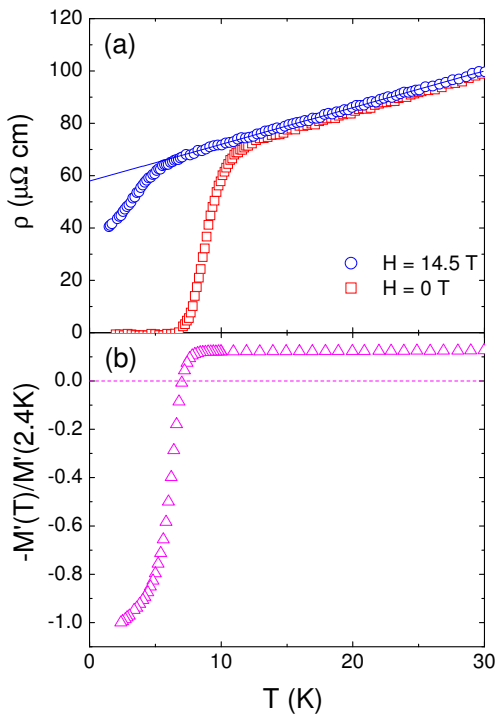}
\caption{(Color online) (a) In-plane resistivity $\rho(T)$ of
FeSe$_x$ single crystal in $H$ = 0 and 14.5 T magnetic fields along
the $c$-axis. The solid line is a linear fit of $\rho(T)$ from 8 to
30 K, which gives the residual resistivity $\rho_0$ = 57.9 $\mu
\Omega$ cm in $H$ = 14.5 T. (b) Normalized ac magnetization.}
\end{figure}

Fig. 1a shows the in-plane resistivity of FeSe$_x$ single crystal in
$H$ = 0 and 14.5 T magnetic fields. The middle point of the
resistive transition is at $T_c$ = 8.8 K in zero field. The 10-90\%
transition width of our crystal is as broad as the powder
sample,\cite{FCHsu} which has been noticed in Ref. 21. Above $T_c$,
$\rho(T)$ manifests a very good linear dependence up to 80 K,
similar to the powder sample.\cite{FCHsu} A linear fit of $\rho(T)$
gives the residual resistivity $\rho_0$ = 57.9 $\mu \Omega$ cm in
$H$ = 14.5 T, which is about 1/4 the value of powder
sample.\cite{FCHsu}

\begin{figure}
\includegraphics[clip,width=7.5cm]{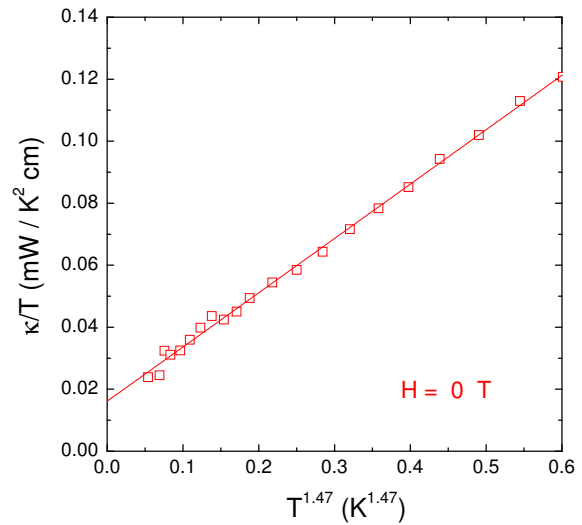}
\caption{(Color online) Temperature dependence of the in-plane
thermal conductivity for FeSe$_x$ single crystal in zero field. The
solid line represents a fit of the data to $\kappa/T = a +
bT^{\alpha-1}$. This gives the residual linear term $\kappa_0/T$ =
16 $\pm$ 2 $\mu$W K$^{-2}$ cm$^{-1}$.}
\end{figure}

To estimate the upper critical field $H_{c2}(0)$ which completely
suppresses the resistive transition, we define $T_c(onset)$ at the
temperature where $\rho(T)$ deviates from the linear dependence, and
get $T_c(onset)$ = 11.9 and 6.3 K for $H$ = 0 and 14.5 T,
respectively. Using the relationship $H_{c2}/H_{c2}(0) = 1 -
(T_c/T_c(0))^2$, we get $H_{c2}(0)$ = 20.1 T. Note that $H_{c2}(0)$
= 16.3 T was estimated for the powder sample, in which $T_c$ was
defined at the middle point of the transition.\cite{FCHsu}

In Fig. 1b, the normalized ac magnetization for FeSe$_x$ single
crystal is plotted. The positive ferromagnetic background has been
attributed to the existence of Fe impurity in the FeSe$_x$ powder
sample.\cite{FCHsu} However, no iron, iron oxide, or iron silicide
impurities were detected in our crystals,\cite{UPatel} therefore the
ferromagnetic background likely results from the magnetic Fe cluster
promoted by Se vacancies.\cite{KWLee}

Fig. 2 shows the temperature dependence of the in-plane thermal
conductivity for FeSe$_x$ in zero field. To extrapolate the residual
linear term $\kappa_0/T$, we fit the data to $\kappa/T = a +
bT^{\alpha-1}$,\cite{Sutherland,SYLi} where $aT$ and $bT^{\alpha}$
represent electronic and phonon contributions, respectively. In Fig.
2, the data from 120 mK to 0.7 K can be fitted (the solid line) and
gives $\kappa_0/T$ = 16 $\pm$ 2 $\mu$W K$^{-2}$ cm$^{-1}$, with
$\alpha$ = 2.47.

Such a value of $\kappa_0/T$ is slightly larger than the
experimental error bar $\pm$ 5 $\mu$W K$^{-2}$ cm$^{-1}$.\cite{SYLi}
However, it is still fairly small, less than 4\% the normal state
Wiedemann-Franz law expectation $\kappa_{N0}/T = L_0/\rho_0$ = 0.423
mW K$^{-2}$ cm$^{-1}$, with $L_0$ the Lorenz number 2.45 $\times
10^{-8}$ W $\Omega$ K$^{-2}$ and $\rho_0$ = 57.9 $\mu \Omega$ cm.
For unconventional superconductors with nodes in the superconducting
gap, a substantial $\kappa_0/T$ in zero field contributed by the
nodal quasiparticles has been found.\cite{Proust,Suzuki} For
example, for overdoped $d$-wave cuprate superconductor Tl2201 with
$T_c$ = 15 K, $\kappa_0/T$ = 1.41 mW K$^{-2}$ cm$^{-1}$, about 36\%
$\kappa_{N0}/T$.\cite{Proust} For $p$-wave superconductor
Sr$_2$RuO$_4$ with $T_c$ = 1.5 K, $\kappa_0/T$ = 17 mW K$^{-2}$
cm$^{-1}$, more than 9\% $\kappa_{N0}/T$ for the best
sample.\cite{Suzuki} We also note that $\kappa_0/T$ in zero field
are all negligible in closely related superconductors
BaNi$_2$As$_2$, Ba$_{1-x}$K$_x$Fe$_2$As$_2$, and
BaFe$_{1.9}$Ni$_{0.1}$As$_2$.\cite{NKurita,XGLuo,LDing} Therefore,
it is unlikely that $\kappa_0/T$ = 16 $\pm$ 2 $\mu$W K$^{-2}$
cm$^{-1}$ in FeSe$_x$ single crystal comes from the nodal
quasiparticles. Since no impurity phases were detected, such a small
$\kappa_0/T$ may simply come from the slight overestimation when
doing extrapolation, due to the lack of experimental data below 120
mK.

Below we turn to the field dependence of $\kappa_0/T$ in FeSe$_x$.
Fig. 3 shows the low-temperature thermal conductivity of FeSe$_x$ in
magnetic fields applied along the $c$-axis ($H$ = 0, 1, 4, 9, and
14.5 T). For $H$ = 1 T, the data is also fitted to $\kappa/T = a +
bT^{\alpha-1}$, and gives $\kappa_0/T$ = 47 $\pm$ 2 $\mu$W K$^{-2}$
cm$^{-1}$, with $\alpha$ = 2.47. For $H$ = 4, 9, and 14.5 T, the
electronic contribution becomes more and more dominant and the data
get less smooth, therefore $\alpha$ is fixed to 2.47 in the phonon
term $bT^\alpha$. From Fig. 3, even higher magnetic field is needed
to increase $\kappa/T$ to its normal state value.

\begin{figure}
\includegraphics[clip,width=7.6cm]{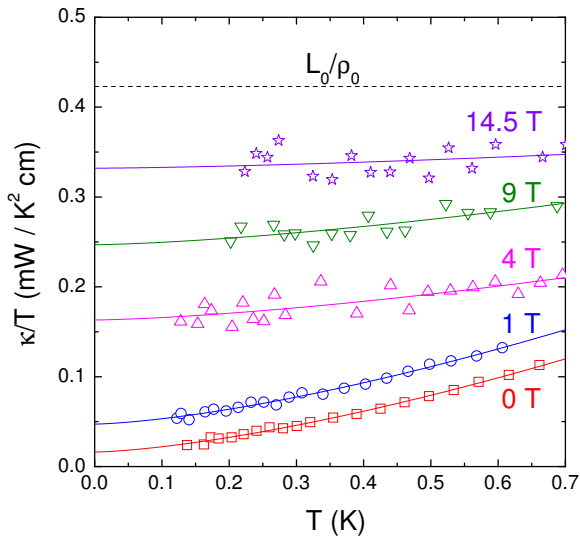}
\caption{(Color online) Low-temperature thermal conductivity of
FeSe$_x$ in magnetic fields applied along the $c$-axis ($H$ = 0, 1,
4, 9, and 14.5 T). The solid lines are $\kappa/T = a +
bT^{\alpha-1}$ fits. For $H$ = 4, 9, and 14.5 T, the electronic
contribution becomes more and more dominant and the data get less
smooth, therefore $\alpha$ is fixed to 2.47. The dashed line is the
normal state Wiedemann-Franz law expectation at $T \rightarrow 0$,
namely $L_0$/$\rho_0$, with $L_0$ the Lorenz number 2.45 $\times
10^{-8}$ W $\Omega$ K$^{-2}$.}
\end{figure}

\begin{figure}
\includegraphics[clip,width=7.5cm]{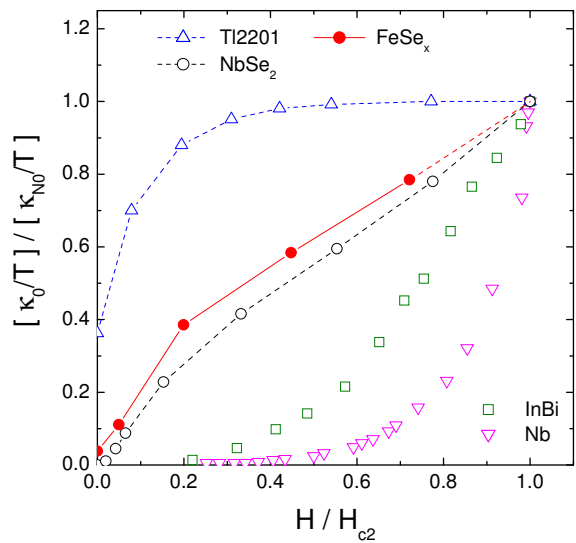}
\caption{(Color online) Normalized residual linear term $\kappa_0/T$
of FeSe$_x$ plotted as a function of $H/H_{c2}$. Similar data of the
clean $s$-wave superconductor Nb,\cite{Lowell} the dirty $s$-wave
superconducting alloy InBi,\cite{Willis} the multi-band $s$-wave
superconductor NbSe$_2$,\cite{Boaknin} and an overdoped sample of
the $d$-wave superconductor Tl-2201 \cite{Proust} are also shown for
comparison.}
\end{figure}

In Fig. 4, we put the normalized $\kappa_0/T(H)$ of FeSe$_x$
together with the clean $s$-wave superconductor Nb,\cite{Lowell} the
dirty $s$-wave superconducting alloy InBi,\cite{Willis} the
multi-band $s$-wave superconductor NbSe$_2$ \cite{Boaknin} and an
overdoped sample of the $d$-wave superconductor
Tl-2201,\cite{Proust} plotted as a function of $H/H_{c2}$. For a
clean (like Nb) or dirty (like InBi) type-II $s$-wave superconductor
with isotropic gap, $\kappa_0/T$ should grow exponentially with
field (above $H_{c_1}$). This usually gives negligible $\kappa_0/T$
for field lower than $H_{c_2}/4$. For the $d$-wave superconductor
Tl-2201, $\kappa_0/T$ increases roughly proportional to $\sqrt{H}$
at low field due to the Volovik effect.\cite{Volovik} By contrast,
for multi-gap superconductors NbSe$_2$ and
MgB$_2$,\cite{Boaknin,AVSologubenko} magnetic field will first
suppress the superconductivity on the Fermi surface with smaller
gap, and give distinct shape of $\kappa_0/T(H)$ curve, as seen in
Fig. 4.

From Fig. 4, the $\kappa_0/T(H)$ of FeSe$_x$ manifests almost
identical behavior as that of multi-gap $s$-wave superconductor
NbSe$_2$. For NbSe$_2$, the shape of $\kappa_0/T(H)$ has been
quantitatively explained by multiband superconductivity, whereby the
gap on the $\Gamma$ band is approximately one third of the gap on
the other two Fermi surfaces.\cite{Boaknin} Therefore, we consider
our data as strong evidence for multi-gap nodeless superconductivity
in FeSe$_x$. Note that in the two-gap $s+s$-wave model to describe
the in-plane penetration depth data, the magnitude of the two gaps
are 1.60 and 0.38 meV, respectively.\cite{RKhasanov} The ratio
between these two gaps is about 4, close to that in NbSe$_2$, thus
supports the multi-gap scenario from our thermal conductivity
results.

So far, there is still no experiment to directly measure the
superconducting gap in Fe$_{1+y}$Te$_{1-x}$Se$_x$ system. Density
functional calculations show that the electronic band structure of
FeS, FeSe, and FeTe are very similar to the FeAs-based
superconductors.\cite{ASubedi} In doped BaFe$_2$As$_2$, multi-gap
nodeless superconductivity has been clearly demonstrated by
angle-resolved photoemission spectroscopy (ARPES)
experiments.\cite{HDing,KNakayama,KTerashima} For hole-doped
Ba$_{0.6}$K$_{0.4}$Fe$_2$As$_2$ ($T_c$ = 37 K), the average gap
values $\Delta(0)$ for the two hole pockets ($\alpha$ and $\beta$)
are 12.5 and 5.5 meV, respectively, while for the electron ($\gamma$
and $\delta$) pockets, the gap value is similar, about 12.5
meV.\cite{HDing,KNakayama} For electron-doped
BaFe$_{1.85}$Co$_{0.15}$As$_2$ ($T_c$ = 25.5 K), the average gap
values $\Delta(0)$ of hole ($\beta$) and electron ($\gamma$ and
$\delta$) pockets are 6.6 and 5.0 meV, respectively
\cite{KTerashima}. The ratio between the large and small gaps is 2.3
for Ba$_{0.6}$K$_{0.4}$Fe$_2$As$_2$. This may explain the rapid
increase of $\kappa_0/T(H)$ at low field in
Ba$_{1-x}$K$_x$Fe$_2$As$_2$,\cite{XGLuo} although magnetic field was
only applied up to 1/4 $H_{c2}$ thus could not see clear multi-gap
character as in our FeSe$_x$ single crystal.

In summary, we have measured the low-temperature thermal
conductivity of iron selenide superconductor FeSe$_x$ to investigate
its superconducting gap structure. A fairly small $\kappa_0/T$ at
zero field and the dramatic field dependence of $\kappa_0/T$ give
strong evidence for multi-gap nodeless superconductivity in
FeSe$_x$. Such a gap structure may be generic for all Fe-based
superconductors. More experiments are needed to distinguish
conventional $s$-wave from the unconventional $s^{\pm}$-wave
superconductivity in this new family of high-$T_c$ superconductors.

This work is supported by the Natural Science Foundation of China,
the Ministry of Science and Technology of China (National Basic
Research Program No:2009CB929203), Program for New Century Excellent
Talents in University, and STCSM of China (No: 08dj1400200 and
08PJ1402100). The work in Northern Illinois University was supported
by the US Department of
Energy Grant No. DE-FG02-06ER46334 and Contract No. DE-AC02-06CH11357. \\

$^*$ E-mail: shiyan$\_$li@fudan.edu.cn

\end{document}